# Interaction-induced interference in the integer quantum Hall effect

I. Sivan,[1] R. Bhattacharyya,[1] H. K. Choi,[1,2,*] M. Heiblum,[1,**] D. E. Feldman,[3] D. Mahalu,[1] and V. Umansky[1]

[1] *Braun Center for Sub-Micron Research, Department of Condensed Matter Physics, Weizmann Institute of Science, Rehovot, Israel 76100*

[2] *Department of Physics, Research Institute of Physics and Chemistry, Chonbuk National University, Jeonju 54896, Korea*

[3] *Department of Physics, Brown University, Providence, Rhode Island 02912, USA*

## ABSTRACT

In recent interference experiments with an electronic Fabry–Perot interferometer (FPI), implemented in the integer quantum Hall effect regime, a flux periodicity of *h*/2*e* was observed at bulk fillings $\nu_B > 2.5$. The halved periodicity was accompanied by an interfering charge $e^* = 2e$, determined by shot noise measurements. Here, we present measurements, demonstrating that, counterintuitively, the coherence and the interference periodicity of the interfering chiral edge channel are solely determined by the coherence and the enclosed flux of the adjacent edge channel. Our results elucidate the important role of the latter and suggest that a neutral chiral edge mode plays a crucial role in the pairing phenomenon. Our findings reveal that the observed pairing of electrons is not a curious isolated phenomenon, but one of many manifestations of unexpected edge physics in the quantum Hall effect regime.

# I. INTRODUCTION

Electron interferometry has played a significant role in studies of the foundations of quantum mechanics. As fabrication techniques improved and novel materials developed, interferometers were realized in numerous mesoscopic systems, such as quantum Hall systems [1–4], nanowires [5], carbon nanotubes [6,7], and graphene [8,9]. Newly developed electronic interferometers revealed unexpected behavior, mostly attributed to the electrons' interaction [10–14].

The quantum Hall effect (QHE) provides particularly attractive settings for interference experiments, since it allows easy control of the chiral edge channels. The sought after interference in the fractional QHE regime - not observed yet - is particularly important because it is expected to reveal fractional statistics of the quasiparticles [15–20]. Yet, the easy-to-access integer QHE still does not cease to surprise us [14].

Here, we elaborate on our recent findings in a 'screened' Fabry–Perot interferometer (FPI) [11,14]. We focus on the interference of the outermost edge channel (closest to the edge), which separates the regions with filling factors $\nu_B = 1$ and $\nu_B = 0$. We discover an important role of the adjacent first-inner channel separating the regions of $\nu_B = 2$ and $\nu_B = 1$. At bulk fillings $\nu_B \geq 2.5$, we observed the following: (*i*) A half-periodicity in the Aharonov-Bohm (AB) flux, i.e., a periodicity of $\phi_0^* = \frac{h}{2e}$; (*ii*) a doubled charge of the interfering particles $e^* \sim 2e$, where $e$ is the electron charge probed via shot noise; and (*iii*) interference quenched when the first-inner edge channel was grounded, while the outermost channel's average current was not affected. None of these effects were observed at bulk fillings $\nu_B < 2.5$, where the observed flux periodicity was $\phi_0 = \frac{h}{e}$ and the interfering charge was $e$.

In the present study we expand on our initial measurements and focus on the important role of the first-inner edge channel. Based on our findings, we suggest that an emerging neutral mode plays a crucial role in the pairing effect.

## II. BACKGROUND

### A. Measurement setup

The FPIs were realized in a high-mobility 2D electron gas (2DEG), embedded in an MBE-grown GaAs/AlGaAs heterostructures, employing a standard fabrication process. The FPIs consisted of two quantum point contacts (QPCs), serving as input and output 'beam splitters'. Ubiquitous Coulomb-dominated physics was suppressed in favor of coherent AB interference in two different implementations of the FPI [Figs. 1a & 1b] [10,14]. In the implementation shown in Fig. 1b, a metallic (Ti–Au) top-gate covered the interior area of the FPI, thus, providing intra-channel screening. In the implementation shown in Fig. 1a, a small grounded Ohmic contact (alloyed Ni\Ge\Au) was placed in the interior of the FPI. The contact allowed draining electrons from the bulk of the FPI (though not from the edges). We used mostly the latter configuration since it allowed adding other features in the interior of the FPI (see below). Unless we explicitly note otherwise, only the outermost edge channel was allowed to interfere (red lines in Figs. 1c & 1d), whereas the inner edge channels (blue and green lines in Figs. 1c & 1d) were trapped inside the FPI. An ac signal (1μV RMS @ 800 kHz) was applied to the source, while the drain voltage (proportional to the transmission of the FPI) was monitored. A cold homemade preamplifier with small signal gain $G$=11 followed by a room-temperature amplifier with gain 200 (NF SA-220F5) were used to amplify the drain's signal.

### B. Interference in the FPI

Aharonov-Bohm interference fringes observed in a small FPI (with area $A_{\text{FPI}} = 2\ \mu\text{m}^2$) at bulk fillings $\nu_B = 2$ and $\nu_B = 3$ are plotted in Fig. 2 as functions of the magnetic field $B$ and the modulation gate voltage $V_{\text{MG}}$. At $\nu_B = 2$, the AB period $\Delta B$ corresponded to an area $\frac{\phi_0}{\Delta B} = 2.05\ \mu\text{m}^2$. However, at $\nu_B = 3$, it corresponded to an 'area' $\frac{\phi_0}{\Delta B} = 4.1\ \mu\text{m}^2$. These two AB frequencies had similar visibilities [14].

One may try to explain this frequency-doubling by assuming that electron trajectories can only wind an even number of times around the FPI; however, this would lead to a considerably lower visibility than observed [14]. Moreover, and most importantly, this would not explain the observed doubling of the interfering charge.

Note that such "pairing" was not observed in a Mach–Zehnder interferometer (MZI) (see Appendix B).

### C. Quasiparticle charge and visibility

As the pairing phenomenon was tied to interference, the charge was measured as a function of the interference visibility, which was suppressed in two ways: (*i*) redirecting the first-inner edge channel to the inner grounded contact, thereby dephasing the FPI (see next chapter) [14]; and (*ii*) increasing the transmission of a single QPC of the FPI. Consequently, the interfering charge $e^*=2e$ measured at a visibility of ~45%, decreased gradually to $e^*=e$ when the visibility was quenched [21]. Yet, as the visibility decreased, the $h/2e$ periodicity remained unchanged. As we argue below, the only way to understand this phenomenon is to assert that single-electron interference is fully dephased in the pairing regime.

## III. ROLE OF THE FIRST-INNER CHANNEL

### A. Grounding the first-inner channel

With a voltage applied to the 'center-QPC' [yellow, Fig. 1d], each edge channel can be reflected separately towards the grounded center-ohmic contact [green, Fig. 1d]. At $\nu_B = 2$, grounding the first-inner channel had no observable effect on the interference of the outermost channel, and the visibility quenched only when the latter was grounded - as expected [Figs. 3a & 3c]. A dramatically different effect was observed at $\nu_B \sim 3$ [Figs. 3b & 3d]. The interference fully quenched when the first-inner channel was directed to ground, thus dephased, with no change of the average current carried by the outermost channel. Evidently, this implies that the coherence of the first-inner channel is indispensable for the pairing phenomenon.

### B. Tuning the AB flux enclosed by the first-inner channel

A different implementation of the FPI, seen in Fig. 4, included an additional island in its interior. This allowed changing the enclosed flux, $BA_{\text{island}}$, by the propagating channels [Fig. 4, green, false color]. Two QPCs were placed in the interior of the FPI: a 'QPC-down' and a 'QPC-up' [Fig. 4, turquoise and purple, respectively]. Pinching QPC-down while keeping QPC-up fully open redirected channels to go around the island. Pinching also QPC-up redirected the channels reflected by QPC-down into the ohmic contact (see Appendix D for a modified configuration of the QPC-down).

An electron in the interferometer accumulates an AB phase $\delta\varphi_{AB} = 2\pi \cdot \frac{(n \cdot A_{FPI} - m \cdot A_{island})\delta B}{\phi_0}$, where $n$ and $m$ represent the numbers of windings that an electron makes around the FPI and the island, respectively. Hence, the two fundamental periods in $B$ are $\phi_0/A_{FPI}$ and $\phi_0/(A_{FPI} - A_{island})$, with the visibility determined by the transmissions of the FPI and the state of QPC-down (with a fully open QPC-up).

At $\nu_B = 2$, the AB frequency and the visibility of the interference are plotted in Figs. 5a & 5b as a function of the voltage applied to QPC-down. A transition in the AB frequency from $\frac{\phi_0}{\Delta B} = A_{FPI} = 12.82$ μm$^2$ to $\frac{\phi_0}{\Delta B} = A_{FPI} - A_{island} = 11.22$ μm$^2$ is observed when QPC-down reflects the outermost edge channel to go around the island [vertical dashed line in Fig. 5a and in Appendix E].

At $\nu_B = 3$, the transition between the two AB frequencies took place, surprisingly, when the first-inner channel was fully reflected to go around the island, while the outermost (interfering) channel was fully transmitted through QPC-down. Here, the AB frequencies switched from $\frac{\phi_0}{\Delta B} = 25.24$ μm$^2$ to $\frac{\phi_0}{\Delta B} = 21.64$ μm$^2$ [see Figs. 5d & 5e and Appendix E]. Note that the innermost (second-inner) channel was never observed to play a direct role in the interference. We believe that its role was to screen the two outer channels from the bulk (that is usually dissipative and electrically noisy).

The experiment was repeated with a modified FPI where QPC-down was replaced by a wider constriction (width ~0.5μm, hereafter named 'wide constriction'). While at $\nu_B = 2$ the behavior did not change, a markedly different behavior was observed at $\nu_B = 3$ [see Figs. 6a & 6b, and in Appendix E]. Again, the AB frequency switched from $\frac{\phi_0}{\Delta B} = 23.2$ μm$^2$ to $\frac{\phi_0}{\Delta B} = 17.8$ μm$^2$ when the wide constriction gradually pinched. However, the two regions, each governed by one of these two frequencies, were separated by a wide transition region with very low visibility. This transition region corresponded to the first-inner channel being fully reflected by the wide constriction, while the outermost edge channel remained unperturbed. In order to explain this finding we propose a model where a neutral quasiparticle is formed due to the interaction between the two outer channels. The neutral mode has a finite transmission even in a pinched QPC constriction but a much smaller transmission in a wide constriction. We elaborate further on this model in the Discussion.

### C. Direct test of the interaction between the two outer edge channels

A screened FPI [such as the one shown in Fig. 1a] was placed in one arm of a Mach-Zehnder interferometer (MZI) [1,22]. The two interferometers allowed interference of the outermost edge channel [Figs. 7, 8b and 8c]. In Figs. 8a and 8d we plot the fast Fourier transform (FFT) of the interference oscillations for different transmissions of the FPI's QPCs (at a constant transmission of the MZI's QPCs).

We first studied the configuration in which the FPI fully transmitted the outermost channel while the inner channels were trapped inside it. At $\nu_B = 2$, a single frequency corresponding to $\frac{\phi_0}{\Delta B} = 14.39$ μm$^2 = A_{\text{MZI}}$ is observed [see the lowest blue trace in Fig. 8a]. At $\nu_B = 3$, two frequency components appear: one corresponds to $A_{MZI}$, and the other corresponds to $\phi_0/\Delta B = A_{\text{MZI}} - A_{\text{FPI}}$ [see lowest blue trace in Fig. 8d]. This occurs in spite of the fact that the outermost edge channel passed freely through the FPI and never encircled its area $A_{\text{FPI}}$.

This phenomenon can be understood as follows. With increasing magnetic field, the enclosed area by the first-inner edge channel, $A_{\text{FPI}}^{(in)}$, decreases slightly in order to maintain a constant flux (and charge) within it [23]. Once this area changes by an area containing a flux quantum, $\delta A_{\text{FPI}}^{(in)} = \phi_0/B$, an electron is added to the first-inner edge channel (rather abruptly), and the area returns to its original size, $\delta A_{\text{FPI}}^{(in)} \to 0$. This process leads to 'breathing-like' periodic modulations of the area, with the frequency $\frac{\phi_0}{\Delta B} = A_{\text{FPI}}$ [23]. Now, due to Coulomb interaction between the outermost and the first-inner edge channels, the area enclosed by the outermost channel breathes as well. This breathing of the area has two implications on the outermost channel's interference in the MZI. It leads to the straightforward modulation of the MZI AB phase, but also to the modulation of the interference visibility as explained in detail in Appendix F. These two effects together give rise to the appearance of the frequency $\phi_0/\Delta B = A_{\text{MZI}} - A_{\text{FPI}}$ even when the outermost edge channel passes freely through the FPI. While this effect should in principle apply also at $\nu_B = 2$, it was not observed. Obviously, such strong inter-mode interaction is unique to the $h/2e$ regime. Note that such a signature of the interaction between the two outer edge channels is not observable in the stand-alone FPI.

## IV. IS PAIRING A SINGLE-PARTICLE PHENOMENON?

With the same MZI-FPI structure, we tuned the FPI's QPCs to partition the outermost edge channel. It is expected that if pairing were a single-particle phenomenon, the transmission phase of the FPI would subtract from the AB phase of the MZI [due to the opposite chirality in the FPI; Fig. 7 and Fig 8c]. Under this hypothesis, the transmission amplitude of the combined MZI-FPI interferometer would be:

$$\tau_{MZI-FPI} = t^2 + r^2 e^{i\varphi_{MZI}} \cdot \tau^2 \sum_{n=0}^{\infty} \left(\rho^2 e^{-i\varphi_{FPI}}\right)^n , \quad (1)$$

where $t$ ($\tau$) and $r$ ($\rho$) are the real transmission and reflection amplitudes of the QPCs of the MZI (FPI), respectively, and $n=0, 1, 2,...$ denotes the number of windings in the FPI. The transmission probability is composed of three oscillatory terms (plus a flux-independent constant):

$$T = |\tau_{MZI-FPI}|^2 = T_{MZI} + T_{FPI} + T_{MZI-FPI} + Const. \quad (2)$$

$$T_{MZI} = \tau^2 C_1 \cos\varphi_{MZI} , \quad (2a)$$

$$T_{FPI} = r^4 \sum_{n=1}^{\infty} D_n \cos(n\varphi_{FPI}) , \quad (2b)$$

$$T_{MZI-FPI} = \sum_{n=1}^{\infty} E_n \cos(\varphi_{MZI} - n\varphi_{FPI}) , \quad (2c)$$

where $T_{MZI}$ and $T_{FPI}$ are the transmission probabilities (up to a constant) of each interferometer independently. The two probabilities oscillate with the frequencies $\frac{\phi_0}{\Delta B} = A_{MZI}$ and $nA_{FPI}$, respectively. The last term $T_{MZI-FPI}$ is due to the interference between the lower path of the MZI and the multiple windings $n$ in the FPI; with each $n$ leading to a different AB frequency, $\frac{\phi_0}{\Delta B} = A_{MZI} - nA_{FPI}$. Naturally, in a single-particle picture, the MZI-FPI structure should show all the frequencies found in Eq. (2), with their relative amplitudes set by the QPCs' transmissions and the coherence length.

At $\nu_B = 2$, as the FPI pinches gradually, the frequency corresponding to $\frac{\phi_0}{\Delta B} = 2.16 \ \mu m^{-2} = A_{FPI}$ appears first, followed by higher harmonics $\frac{\phi_0}{\Delta B} = 2A_{FPI}, 3A_{FPI}, ...,$ as well as $\frac{\phi_0}{\Delta B} = A_{MZI} - A_{FPI}, A_{MZI} - 2A_{FPI}, A_{MZI} - 3A_{FPI}$, as expected [Fig. 8a]. At $\nu_B = 3$, on the other hand, the results do not agree with the single-particle picture [Fig. 8d]. First, we find the doubled frequency $\frac{\phi_0}{\Delta B} = 2A_{FPI}$ and its first harmonic $\frac{\phi_0}{\Delta B} = 4A_{FPI}$, due to paired electrons interference in the FPI. In addition, the frequency component $A_{MZI} - A_{FPI}$ results from the area modulations of the first-inner channel's (as explained above and in Appendix F). Then, as the FPI pinches further [towards the highest graph in Fig. 8d], the first-inner channel gets more confined, leading to longer life time of each electron state, and thus to abrupt area fluctuations and the higher harmonic $A_{\text{MZI}} - 2A_{\text{FPI}}$. Being nearly independent of further pinching of the FPI, the latter two frequencies clearly do not result from single-particle interference in the FPI. Indeed, the paired electrons giving rise to interference in the FPI do not interfere with the electrons traversing the lower arm of the MZI.

## V. DISCUSSION

From its discovery, the QHE has been an extraordinarily rich and exciting field of research. Some of its developments led to novel ideas and advances well beyond QHE physics, such as the discovery of topological insulators [24, 25] and the discovery of the connection between QHE physics and knot theory [26]. The field of the fractional QHE is rich in puzzles, such as the nature of the 5/2 liquid [27]. The integer QHE may seem to be a simpler phenomenon, yet the observed pairing constitutes an open puzzle in the field.

An obvious question is whether the *pairing* phenomenon is an isolated puzzle or part of a broader set of phenomena. With this question in mind, we examined several interferometers beyond the simplest FPI setup. We considered a set of modified FPIs, an MZI that can be smoothly converted to an FPI (Appendix B), and an FPI inserted in one of the two arms of an MZI. The most significant observation was the importance of the first-inner edge channel, which determines the interference periodicity and its coherence.

While not having a detailed model for the observed pairing phenomenon, one can envision the following scenario, which may explain the dephasing process of the paired particles, but not the exact mechanism that leads to the formation of a coherent 2*e* interfering quasiparticle. Let us look at the following two configurations, at $\nu_B = 3$: (a) the first-inner channel encloses a different AB area than that of the outermost channel [Figs. 5d & 6]; (b) the grounding of the first-inner channel leads to dephasing of the outermost channel [Fig. 3b)]. Inter-channel correlations between the two outer channels can be modeled in the following manner [Fig. 9]: Particle A with charge –*q* moves along the outermost (interfering) channel and induces a screening charge distribution in the first-inner channel. We depict the distribution as a combination of two charged objects B and C, with B carrying charge +*q*, compensating the charge –*q* of C [Fig. 9b]. Having reached the narrow center-QPC constriction, the neutral excitation A–B, which sees a lower barrier than that for the charged excitation, crosses the pinched QPC [Fig. 9c]. The neutral excitation accumulates a null AB phase in its path. Hence, the total AB phase is determined by particle C that follows the inner channel trajectory, diverted to go around the island [Fig. 9c]. This accounts for the experimentally observed dominance of the inner-first channel in terms of the interference frequency and coherence. On the other hand, replacing the narrow center-QPC constriction with a wide constriction, which forces the neutral object B-C to go around the island, quenches the interference altogether. Apparently, the breakdown of the A-B neutral excitation followed by its recreation on the other side of the wide-constriction may introduce an arbitrary phase to the outermost channel, thus resulting in dephasing.

A possible dephasing mechanism originates from the faster propagation of charged excitations in comparison with neutral modes. In a narrow constriction, the neutral object A-B takes a short path. The charged particle C takes a long path around the island. Since C is faster, it catches up with A-B. In the wide-constriction geometry, the neutral object B-C is left behind the charged object A and cannot catch up. Thus, A, B, and C cannot recombine.

Our results demonstrate that a complete explanation of the *pairing* phenomenon should satisfy a considerable number of experimental constraints and involve nontrivial interaction physics. While electron interactions are crucial in the FQHE, the simplest model of IQHE neglects interactions. Yet, the observed pairing phenomenon demonstrates that interactions lead to qualitatively new physics even in the integer QHE regime.

## ACKNOWLEDGMENTS

M. H. acknowledges the partial support of the Israeli Science Foundation, Grant No. ISF-459/16 (ISF), the Minerva foundation, the US–Israel Bi-National Science Foundation (BSF), the German Israeli Foundation (GIF), and the European Research Council under the European Community's Seventh Framework Program (FP7/2007-2013)/ERC Grant Agreement No. 227716. H. K. C. was supported by research funds for newly appointed professors of Chonbuk National University in 2016, and Korea NRF (Grant No. 2017R1C1B3004301 and No. 2018R1A5A1010092). D. E. F.

acknowledges the hospitality of the Weizmann Institute. D. E. F's research was supported in part by the National Science Foundation under Grant No. DMR-1607451.

* hkchoi@jbnu.ac.kr

** moty.heiblum@weizmann.ac.il

# Figures

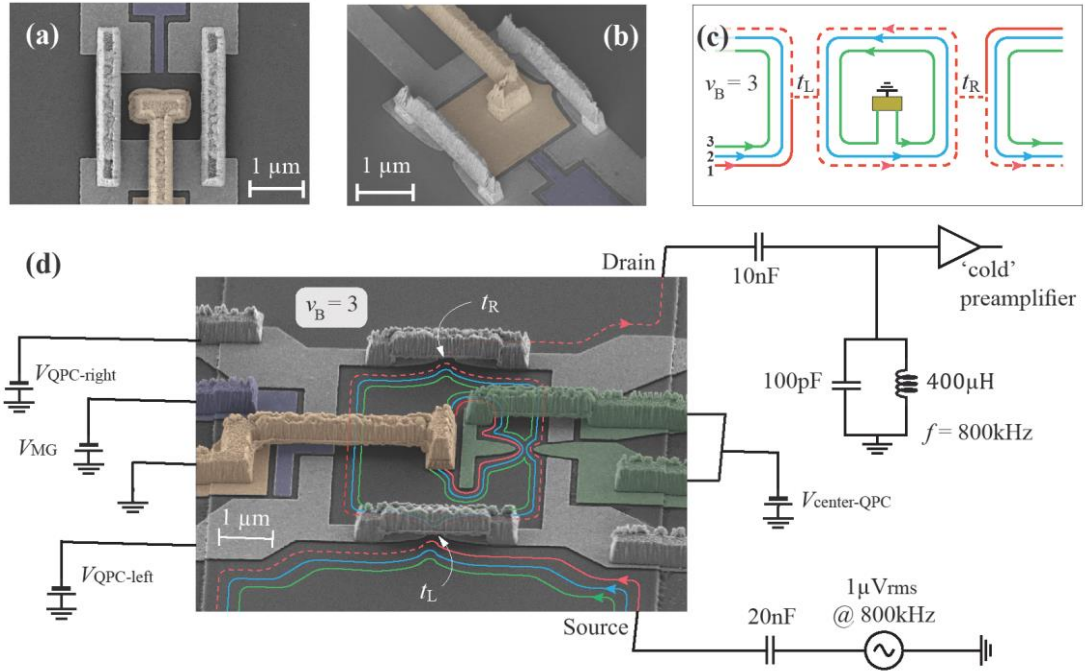

FIG. 1. SEM images of Fabry–Perot interferometers and illustrations of the electronic setup and the chiral edge channels. (a) SEM image of a Fabry–Perot interferometer with a grounded ohmic contact in its center (gold, false color) aimed at suppressing Coulomb blockade physics; such devices show AB interference. (b) SEM image of an FPI covered with a grounded top-gate (gold, false color), which is also aimed at suppressing Coulomb blockade physics. (c) Illustration of the chiral edge channels in our setup at $\nu_B=3$. Red, blue, and green lines represent the outermost, first-, and second-inner edge channels, respectively. Dashed lines represent current that is partitioned by a QPC. In the case presented, the outermost edge channel is partitioned by both QPCs which form the FPI. The ohmic contact is illustrated in the middle of the setup by a yellow box, and the third edge channel is reflected into it by the center-QPC, as seen in (d). (d) SEM image of a 12 μm$^2$ FPI with a center Ohmic contact (gold, false color) and an additional center-QPC (green, false color) placed along the FPI's edge. An illustration of the edge channels, similar to the one in (c), is provided for $\nu_B$~3 with the innermost edge channel reflected by center-QPC into the center ohmic contact. "Cold" edges, originating from the ground, are not plotted for simplicity, and arrows represent the current's chirality. The current is impinged on the device from the source side, partitioned at the two QPCs, and probed at the drain side with a cold amplifier. All three devices [a, b and d] consist of two quantum point contacts, each with a transmission coefficient controlled by the voltage applied to it [$V_{QPC-left}$ and $V_{QPC-right}$ are applied to QPC-left and QPC-right, respectively; see (d)]. A charged "modulation gate" (MG) allowed varying the FPI's area (in all three devices; dark blue, false color).

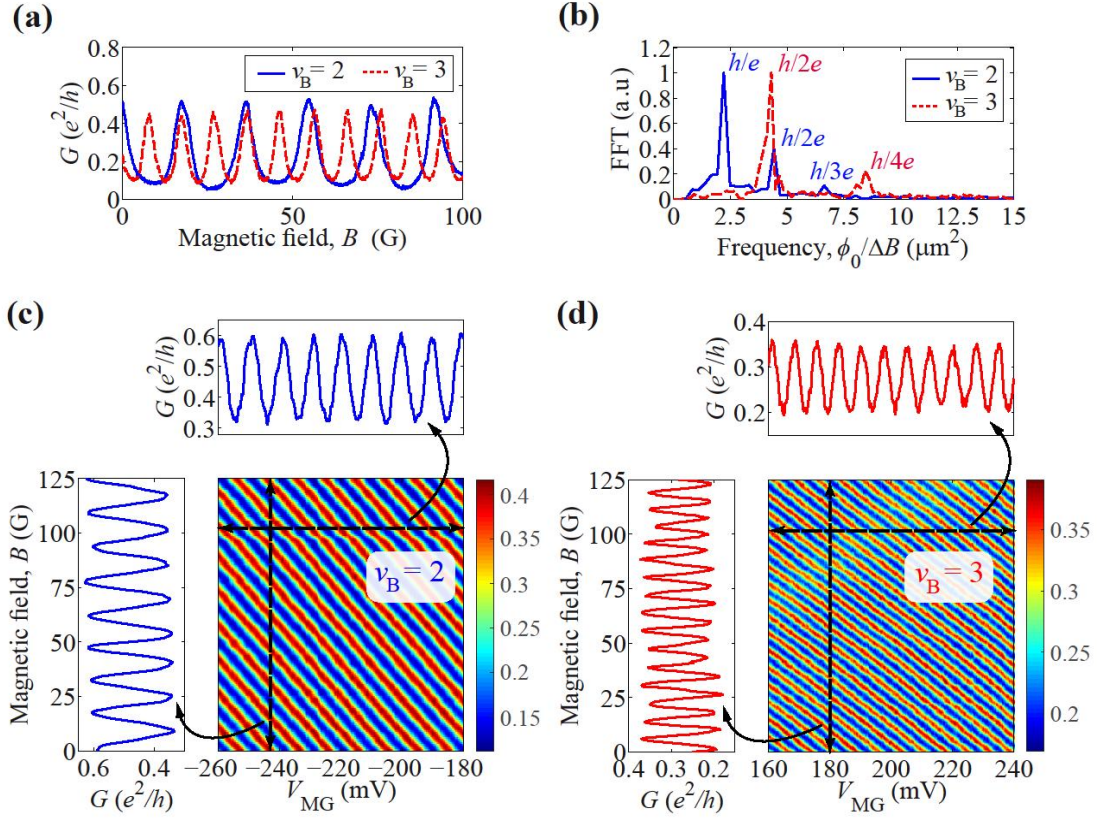

FIG. 2. Aharonov-Bohm interference in the *h/e* and *h/2e* regimes measured with a 2 µm² FPI with a grounded center Ohmic contact. (a) Characteristic AB oscillations with respect to the magnetic field *B* in the two regimes. (b) Corresponding Fourier transforms; clearly, the second harmonic of the *h/e* periodicity coincides with the first harmonic of the *h/2e* periodicity. (c, d) Conductance *G* of the FPI vs both the magnetic field *B* and the modulation-gate voltage $V_{MG}$ in the *h/e* regime (c) and in the *h/2e* regime (d), measured at *B*=5.2T ($v_B$~2) and *B*=3T ($v_B$~3), respectively.

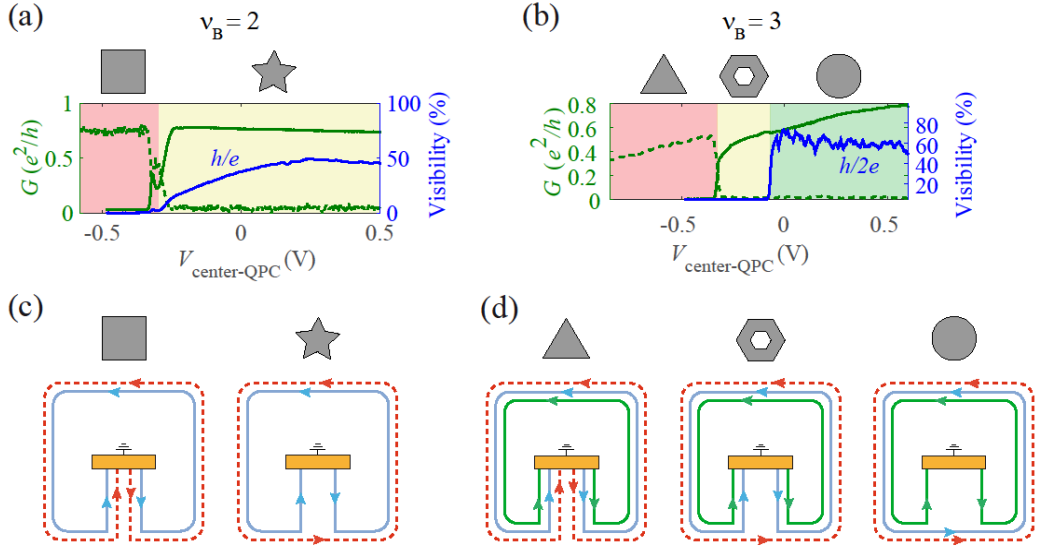

FIG. 3. Grounding edge channels selectively and the effect on interference visibility and conductance. (a, b) Conductance between source and drain through the FPI (green, solid), conductance between source and the center ohmic contact (green, dashed, probed by measuring the voltage across a 1 kΩ resistor bleeding the current in the center contact to ground) and the visibility of the AB interference of the outermost edge channel (blue). All three graphs are plotted as functions of the center-QPC voltage $V_{\text{center-QPC}}$. The results at $\nu_B = 2$ are shown in (a), and the results at $\nu_B = 3$ are shown in (b). Regions that differ by the number of fully transmitted channels at center-QPC are marked with different background colors. These regions are illustrated in (c) and (d) for (a) and (b), respectively.

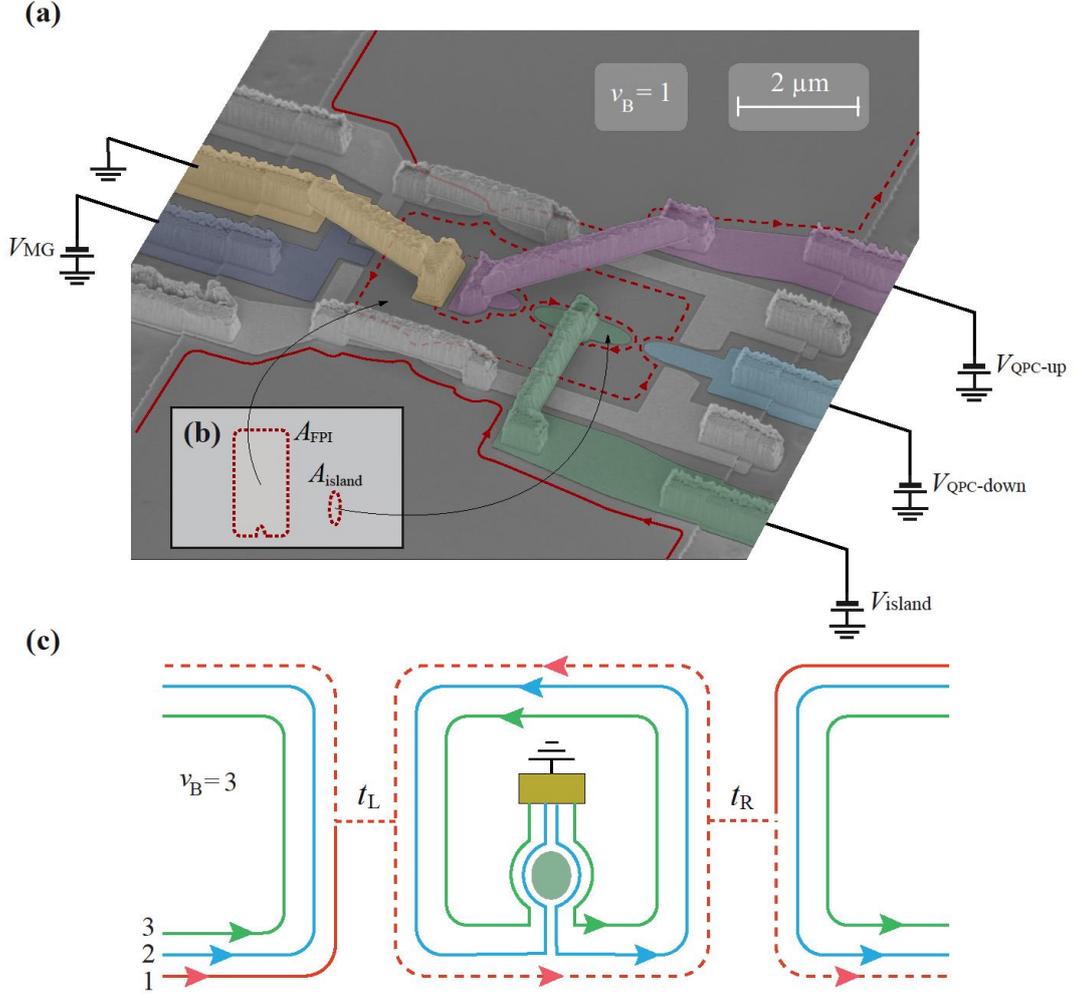

FIG. 4. SEM image of a Fabry-Perot interferometer with an "island" gate in its center. (a) The FPI shown here is similar to the one in Fig. 1d, formed by two QPCs, with a modulation-gate (dark blue, false color) and a center Ohmic contact (gold, false color). An additional gate, biased with $V_{\text{island}}$, is placed in the bulk of the interferometer (green, false color). Together with a second gate, placed along the interferometer's edge (light blue, false color) and biased with $V_{\text{QPC-down}}$, it forms QPC-down, which allows us to reflect edge channels to the island one by one. With a third gate, placed next to the center Ohmic contact (purple, false color) and biased with $V_{\text{QPC-up}}$, QPC-up is formed. QPC-up allows us to reflect edge channels from the island into the grounded center Ohmic contact. We denote the area of the interferometer as $A_{\text{FP}}$ and that of the island as $A_{\text{island}}$. The two areas are marked on top of the image with a dashed red line. (b) Inset: a rescaled version of the two areas, $A_{\text{FPI}}$ & $A_{\text{island}}$, for clarity. (c) An illustration of the interferometer at $\nu_B = 3$ with the outermost edge channel interfering. Edge channels are denoted by colored lines. Dashed lines represent partitioned current. The island gate is denoted by a green ellipse within the FPI's perimeter, and is encircled by the first-inner edge channel. The innermost edge channel is reflected to the island by QPC-down, and then reflected by QPC-up from the island into the center Ohmic contact.

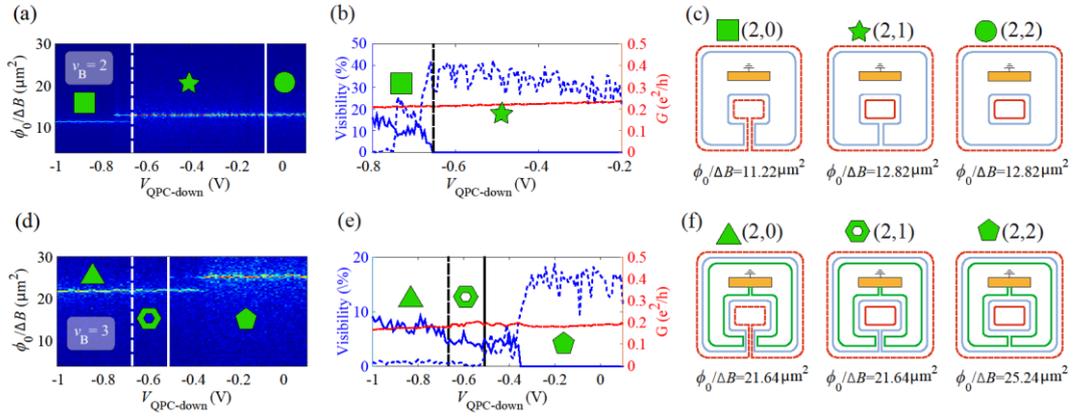

FIG. 5. Evolution of the interference area as QPC-down is pinched, measured with the device in Fig. 4a. (a) Measurements at $\nu_B = 2$. Each vertical line is the fast Fourier transform of the AB interference oscillations measured as a function of the magnetic field $\delta B$ for a particular value of $V_{\text{QPC-down}}$ that determines the transmission of QPC-down. QPC-up is maintained fully open at all times. (b) Conductance (extracted from the AB interference, red) and visibility (extracted from (a), blue) are plotted as functions of $V_{\text{QPC-down}}$. Two visibilities are shown for the two different frequencies seen in (a): $\phi_0/B = 11.22$, marked with a full blue line, and $\phi_0/B = 12.82$ marked with a dashed blue line. (c) Illustrations of different configurations with QPC-down reflecting both edge channels (left), only the inner edge channel (middle) and none of the edge channels (right). Above each illustration we denote the numbers of fully transmitted channels in QPC-down, $\nu_{down}$, and in QPC-up, $\nu_{up}$, in a compact form as $(\nu_{up}, \nu_{down})$. (d–f) Similar measurements and illustrations to those in (a), (b), and (c), respectively, only at $\nu_B = 3$ in the $h/2e$ regime.

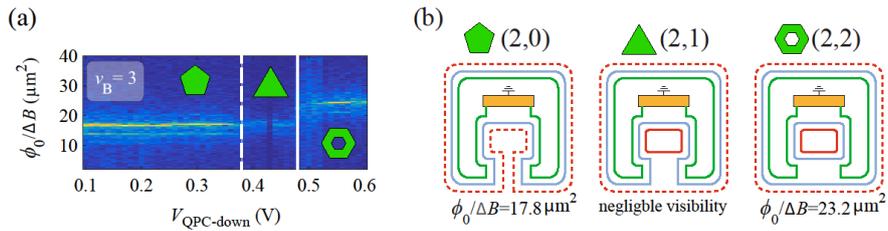

FIG. 6. Evolution of the interference area as QPC-down is pinched, measured with the device from Appendix D at $\nu_B = 3$. (a) Similar measurement to those in Fig. 5d, taken this time with the device that has an wide constriction between the island and the edge of the FPI (replacing QPC-down; see details in Appendix D). (b) Illustrations of the different configurations, similar to those presented in Fig. 5f.

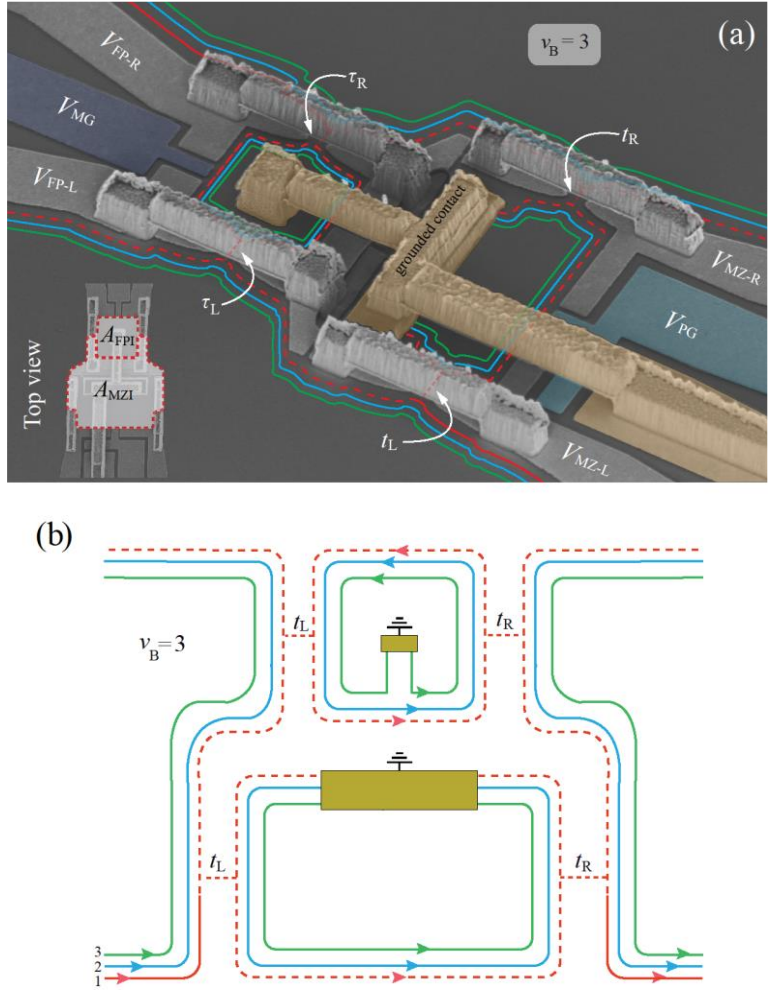

FIG. 7. A Fabry–Perot interferometer on one of the arms of a Mach–Zehnder device. (a) SEM image of the device with an illustration of the chiral edge channels at $\nu_B = 3$. The applied voltages are marked on top of the gates. The modulation gate ($V_{MG}$, dark blue, false color) is utilized in order to modulate the FPI's area. The plunger gate ($V_{PG}$, light blue, false color) is utilized to vary the MZI's area. The edge channels are denoted by the colored lines following the edge profile. Dashed lines stand for edge channels that are partitioned at the QPCs. Both Ohmic contacts (gold, false color) in the center of the FPI and on the edge of the MZI are grounded with a common ground. We denote the area of the MZI as $A_{MZI}$ and that of the FPI as $A_{FPI}$, as seen in the top-view SEM image of the device in the inset. (b) An illustration of the device at $\nu_B = 3$ with the outermost edge channel interfering in both the FPI and the MZI. Edge channels are denoted by colored lines. Partitioned current is denoted by dashed lines.

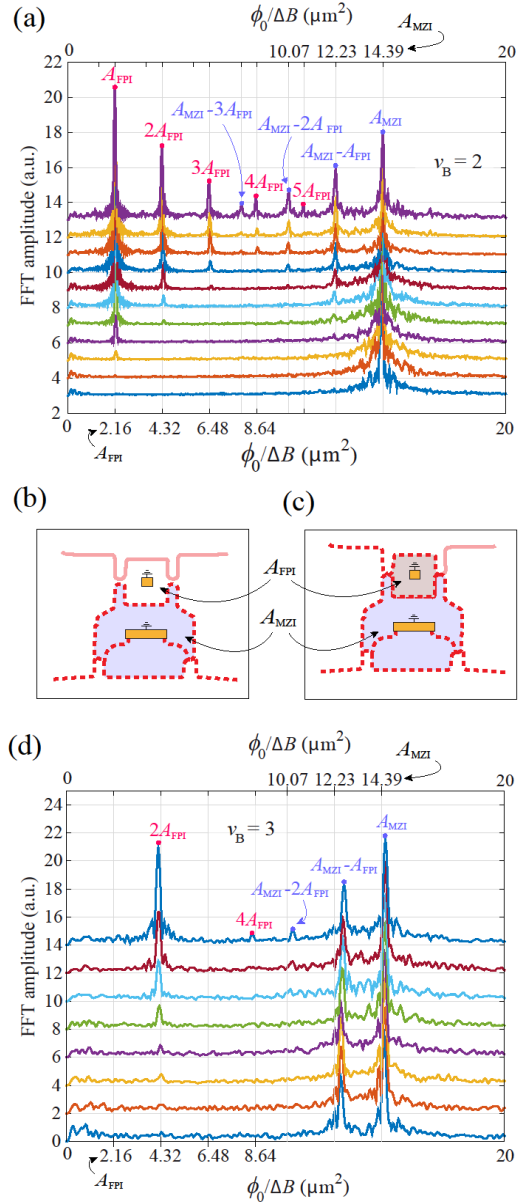

FIG. 8. Dependence to the conductance oscillations' frequencies on the transmission of the QPCs of the device seen in Fig. 7. (a) We set both the MZI's and FPI's QPCs to partition the outermost edge channel at $\nu_B = 2$. While keeping the MZI's QPCs fixed, we varied the transmission of the FPI's QPCs over a wide range. For each value of the FPI QPCs' transmission, we measure interference oscillations as a function of the magnetic field $\delta B$ and plot its fast Fourier transform. This is done starting from the transmission $|t|^2 = 1$ (lowest line, blue) to a considerably low transmission (highest line, purple). Both the x-axes above and below the graph denote the frequency in units of μm². The lower axis shows the multiples of the FPI area ($A_{FPI}, 2A_{FPI}, 3A_{FPI}, ...$), while the upper axis indicates the combinations of the areas of the MZI and FPI ($A_{MZI}, A_{MZI} - A_{FPI}, A_{MZI} - 2A_{FPI}, A_{MZI} - 3A_{FPI}, ...$). (b) An illustration of the device with the QPCs of the MZI partitioning the outermost edge channel, while the QPCs of the FPI are maintained fully open. (c) An illustration of the device with all QPCs partitioning the outermost edge channel. All graphs in (a), except the lowest two, were obtained in that configuration. (d) Similar measurements as shown in (a) at $\nu_B = 3$ in the $h/2e$ regime.

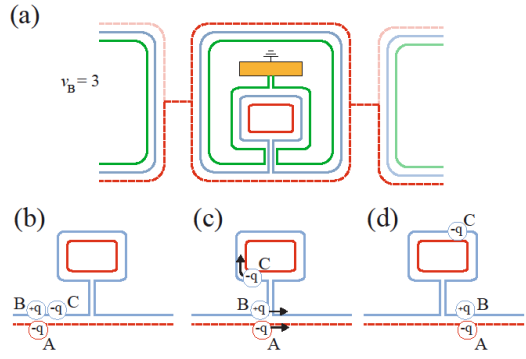

FIG. 9. Schematic representation of the neutral mode tunneling through QPC-down. (a) Illustration of the device shown in Fig. 4a at $\nu_B = 3$. The semi-transparent lines represent channels at ground or 'floating' channels. The dashed lines represent partitioned current. The configuration illustrated here is the same as the one marked with a hexagon in Fig. 5f. (b) Negative charge *A* travels along the chiral outermost channel toward QPC-down. *B* represents a positive screening charge and *C* is a compensating negative charge on the neutral inner channel. (c, d) Dipole *A-B* tunnels through QPC-down, while *C* continues along the inner channel around the island.

# APPENDIX A: NON-INTERACTING FABRY-PEROT INTERFEROMETER

The theory of a non-interacting FPI is well established. The AB phase is given by $\delta\varphi_{AB} = 2\pi \frac{\delta\phi_{FPI}}{\phi_0^*}$ with $\delta\phi_{FPI} = A\delta B + B\delta A$, where $A$ is the area of the device and $B$ is the magnetic field. This phase determines the transmission amplitude $\tau_{FPI}$,

$$\tau_{FPI} = t^2 \sum_{n=0}^{\infty} \left(r^2 e^{i\delta\varphi_{AB}}\right)^n = \frac{t^2}{1 - r^2 e^{i\delta\varphi_{AB}}}, \qquad (A1)$$

where $t = t_L = t_R$ and $r = r_L = r_R$ are the transmission and reflection amplitudes of the left and right QPCs. To first order in $|r|^2 \ll 1$, the transmission coefficient is $T_{FPI} = |\tau_{FPI}|^2 = |t^2(1 + r^2 e^{i\delta\varphi_{AB}})|^2$. This picture holds for the $h/e$ regime at $\nu_B < 2.5$.

# APPENDIX B: A COMBINED MACH-ZEHNDER-FAPRY-PEROT DEVICE

A pairing phenomenon was never observed in an MZI. Indeed, an MZI is topologically different from an FPI. An MZI contains a grounded drain inside the interferometer. This does not allow inner edge channels to be confined inside the device, as they are grounded. Nevertheless, it is possible to construct a structure that allows a continuous transition between an MZI and an FPI [Fig. 10]. Here, two additional QPCs are added to the MZI (a QPC-middle, marked blue, and a QPC-top, marked green). When the two QPCs are fully open, the device functions as an MZI with the area $A_{MZ}$; however, when the two QPCs are pinched, it functions as an FPI with the area $A_{FP}$. We study the evolution during a gradual transition from one interferometer to the other in the pairing regime.

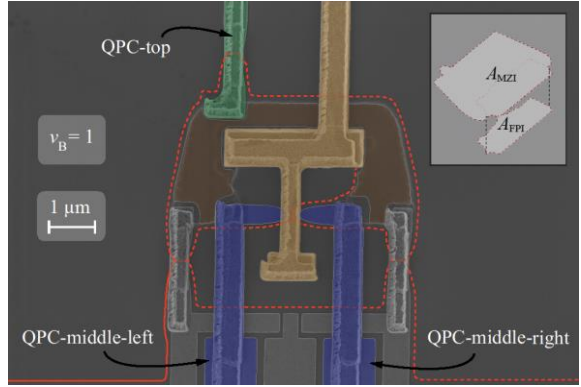

FIG. 10. SEM image of a combined Mach-Zehnder and Fabry-Perot device. The device allows a transition from operating as an MZI to operating as an FPI. Edge channels are depicted for the bulk filling factor $\nu_B = 1$ (red lines) for simplicity. At higher fillings, additional channels appear, as shown in Fig. 1. The partitioned current is denoted by dashed lines. There are two Ohmic contacts connected via an air-bridge (gold, false color). The contacts are grounded together via a second air bridge. The device can be regarded as an MZI with two additional QPCs: QPC-middle (blue, false color) and QPC-top (green, false color). When the two QPCs are fully open, the device functions as an MZI with the area $A_{\text{MZI}}$; however, when the two QPCs are fully pinched, it functions as an FPI with the area $A_{\text{FPI}}$. It can function as a combination of MZI & FPI when QPC-top is fully open, but QPC-middle is set to partially pinch the outermost edge channel. The gradual transition from one interferometer to the other is studied in Fig. 11.

The AB frequency of the MZI configuration at $\nu_B = 2$ corresponds to the area $A_{\text{MZI}} = \frac{\phi_0}{\Delta B} = 21.3 \, \mu m^2$ [Fig. 11a, lowest plot, light blue]. A transition to an FPI configuration takes place when QPC-top fully transmits the outer edge channel and QPC-middle gradually pinches (toward the upper plots in red). In that case, the big Ohmic inside the MZI serves as a common drain for both MZI and FPI configurations. Therefore, the FPI is formed by QPC-middle, QPC-left and QPC- right, while these last two also form the MZI. Additional prominent frequency components appear, which correspond to $A_{\text{FPI}} = \phi_0/\Delta B = 5.7 \, \mu m^2$, as well as to the sum and difference of the two areas: 27 µm² and 15.3 µm².

At $\nu_B = 3$, the frequency in the MZI configuration [Fig. 11b, lowest plot, light blue] is nearly identical to that in Fig. 11a. Then, as QPC-middle pinches, the doubled frequency of the FPI, corresponding to $2A_{\text{FPI}} = 11.4 \, \mu m^2$, emerges. In contrast to $\nu_B = 2$, when the FPI forms (middle graph, purple) and the $2A_{\text{FPI}}$ component increases, the $A_{\text{MZI}} = 21.3 \, \mu m^2$ component decreases substantially. This suggests the dephasing of the single particle frequency component corresponding to $A_{\text{MZI}}$ when that of the paired electrons in the FPI increases. Moreover, while at $\nu_B = 2$ the frequencies $\frac{\phi_0}{\Delta B} = A_{\text{MZI}} \pm A_{\text{FPI}}$ appear, at $\nu_B = 3$ only a weak high band $A_{\text{MZI}} + A_{\text{FPI}}$ is apparent. Clearly, no frequency suggestive of pairing, $\frac{\phi_0}{\Delta B} = 2A_{\text{MZI}}$, is observed.

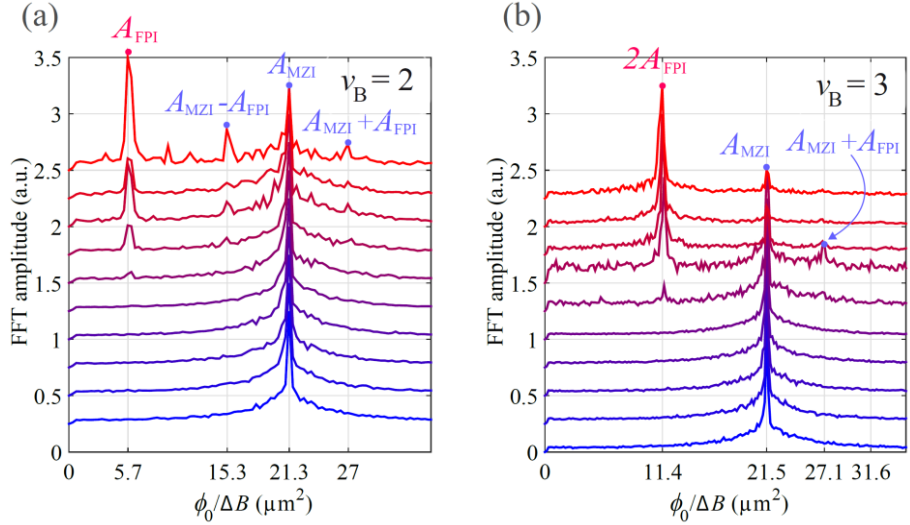

FIG. 11. AB interference frequencies for different configurations of the device shown in Fig. 10. (a) Measurements at $v_B = 2$. Each vertical line is the fast Fourier transforms (FFT) for a particular value of the FPI's transmission. Peaks represent dominant frequencies of the interference fringes. The frequencies' values in terms of the physical areas $A_{\text{MZI}}$ & $A_{\text{FPI}}$ are marked on top of the graph. Measurements were performed from full transmission of the outermost edge channel (lowest graph, blue) to its very low transmission, $t \ll 1$, (highest graph, red). The MZI's QPCs partition the outermost edge channel and were kept constant for all measurements. (b) Similar measurements to (a), but at $v_B = 3$.

# APPENDIX C: EVOLUTION OF THE 2$e$ CHARGE ACCORDING TO THE VISIBILITY OF THE FABRY-PEROT INTERFEROMETER

We studied the dependence of the quasiparticle charge on the visibility of the FPI at $v_B = 3$. In this experiment, the QPC-left of the FPI is set to be at a fixed transmission, $t_L$, and the transmission of the QPC-right, $t_R$, is varied from 0 to 1 for the outermost edge channel. The quasiparticle charge is measured via shot noise as the visibility of the interference changes.

As the doubling of the charge, extracted from shot-noise measurements [28–31], cannot be easily understood, the actual interfering charge was verified by employing a few versions of the standard non-interacting expression for the spectral density of the shot noise. An expression that takes into account "bunching" of charge in the partitioning process, where the incoming quasiparticles' charge is smaller than that of the partitioned ones, is relevant to this phenomenon [32,33]: $S_{shot\ noise} = 2e^* I t \left(1 - t\frac{e}{e^*}\right) L(e^*, V, T)$, where $I$ is the impinging current at the FPI, $t$ is the total transmission through FPI, $e^*$ is the charge of the partitioned quasiparticles, $V$ is the bias voltage, $T$ is the temperature, and $L(e^*, V, T)$ is the temperature-dependent Langevin function $L(e^*, V, T) = \coth\left(\frac{e^*V}{2k_BT}\right) - \frac{2k_BT}{e^*V}$ [32].

Fig. 12 shows that the interfering charge, $e^*$, becomes close to $2e$ when the visibility is highest. On the other hand, it is $e^*=e$, when the visibility goes to zero, and $e \leq e^* < 2e$ elsewhere. The charge was not measured for very small transmissions through the FPI as the shot-noise curve fitting becomes unreliable in such cases. Intermediate values of the charge between $e$ and $2e$ suggest the coexistence of quasiparticles with the charges $2e$ and $e$; yet, no $h/e$ interference was found. One must conclude that single-particle interference is dephased in this regime.

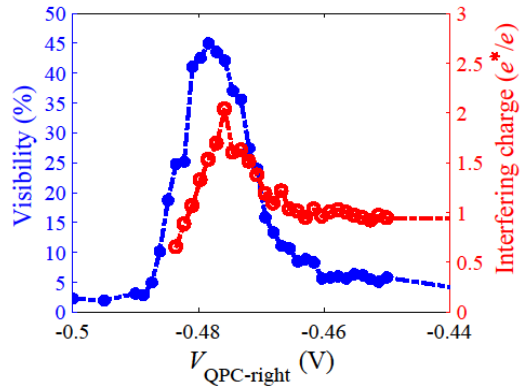

FIG. 12. The dependence of the quasiparticle charge on the visibility as one of the FPI's QPCs is gradually opened at $\nu_B = 3$. The blue curve shows the evolution of the interference visibility as a function of $t_R$ (in terms of the voltage applied to QPC-right) for a fixed $t_L$. With the same QPC transmissions, the quasiparticle charge in the FPI is measured via shot noise (red). Notably, the evolution of the interfering charge follows the evolution of the visibility.

# APPENDIX D: EMPLOYING AN EXTENDED CONSTRICTION

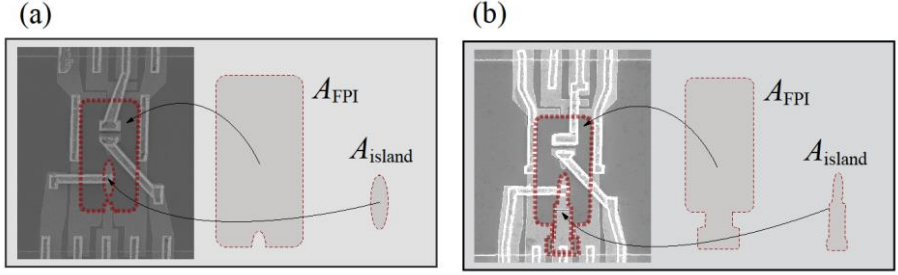

FIG. 13. Top-view SEM images of a Fabry–Perot interferometer with an "island" in its center. (a) Top-view SEM image of the device shown in Fig. 4 alongside illustrations of $A_{\text{FPI}}$, the FPI's area, and $A_{\text{island}}$, the island's area. (b) Top-view SEM image of a device similar to the one in (a) but with a wide constriction between the island and the edge of the FPI. As discussed in the main text, the effect of reflecting the edge channels one by one from the interferometer's edge to the island differs in these two device.

# APPENDIX E: THE EFFECT OF SELECTIVE REFLECTION OF EDGE CHANNELS INTO THE GROUND

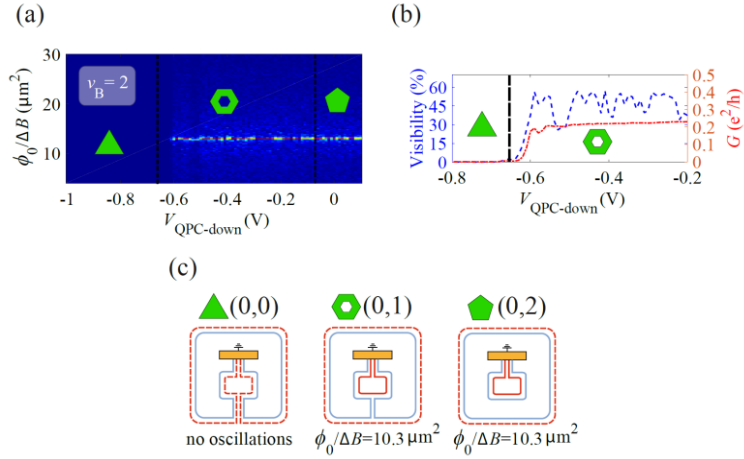

FIG. 14. Evolution of the interference as QPC-down is pinched, measured with the device shown in Fig. 4a at $\nu_B = 2$ with QPC-up fully closed. (a) 2D plot of the fast Fourier transform of the AB interference oscillations as a function of the magnetic field $\delta B$ throughout a wide range of values of $V_{\text{QPC-down}}$. Each vertical line is the FFT for a particular value of $V_{\text{QPC-down}}$ and they determines the transmission of QPC-down. The voltages at which the different edge-channels are reflected are marked on top of the graph by vertical dashed lines. The left vertical line (around $V_{\text{QPC-down}} = -0.65$) mark the voltage at which the outermost edge channel is reflected, and the right vertical line (around $V_{\text{QPC-down}} = -0.1$) mark the voltage at which the inner edge channel is reflected. (b) Conductance (extracted from the AB interference) and visibility (extracted from (a)) are plotted as functions of $V_{\text{QPC-down}}$. (c) Illustrations of different configurations with QPC-down reflecting both edge channels (left), only the outermost edge channel

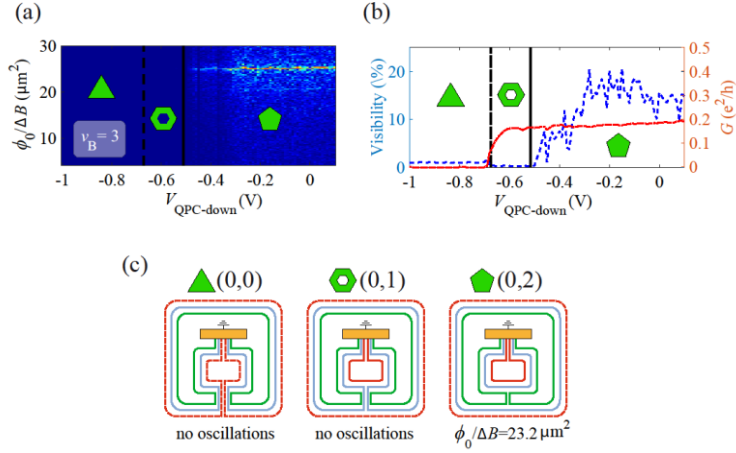

FIG. 15. Evolution of the interference area as QPC-down is pinched, employing the device shown in Fig. 4a at $\nu_B = 3$ with QPC-up fully closed. (a) Measurements at $\nu_B = 3$. 2D plot of the fast Fourier transform of the AB interference oscillations with respect to the magnetic field $\delta B$ as a function of $V_{\text{QPC-down}}$. Each vertical line is the FFT for a particular value of $V_{\text{QPC-down}}$ determines the transmission of QPC-down. The voltages at which outermost different edge-channels are reflected are marked on top of the graph by vertical lines. The left vertical line (around $V_{\text{QPC-down}} = -0.7$) mark the voltage at which the outermost edge channel is reflected, and the right vertical line (around $V_{\text{QPC-down}} = -0.5$) mark the voltage at which the first-inner edge channel is reflected. (b) Conductance (extracted from the AB interference, red) and visibility [extracted from (a), blue] are plotted as functions of $V_{\text{QPC-down}}$ for a fully closed QPC-up. (c) Illustrations of different configurations with QPC-down reflecting both edge channels (left), only the outermost edge channel (middle), and none of the edge channels (right). Above each illustration, we show as $(\nu_{up}, \nu_{down})$ the numbers of fully transmitted channels in QPC-down, $\nu_{down}$, and in QPC-up, $\nu_{up}$.

# APPENDIX F: SIMULATION FOR THE CONDUCTANCE OF THE MACH-ZEHNDER-FABRY-PEROT DEVICE

We analyze here the conductance oscillations measured with the device shown in Fig. 7. When the FPI fully transmits the outermost channel that arrives from the first (left) QPC of the MZI, while the first-inner channel is confined, a frequency corresponding to $A_{\text{MZI}} - A_{\text{FPI}}$ is observed [Fig. 8d]. This frequency component persists as the FPI is pinched. This frequency can be derived from the Friedel sum rule [24], which relates the scattering phase with the electron occupancy. In our device, an electron added to the confined first-inner edge channel is screened by electrons in the outermost edge channel. While electrons in the outermost channel pass through the FPI without backscattering, they do acquire a scattering phase that depends on the number of electrons in the first-inner edge channel. For every additional electron in the confined first-inner edge channel, a $\pi$ is added to the phase of the electrons in the outermost edge channel. Thus, at these degeneracy points, the MZI is dephased [22]. In Fig. 16 we show a simulation of the interference in this configuration while taking this effect into account. As the line-shapes of the MZI's visibility dips (for the outermost edge channel's interference) follow those of the FPI's conductance peaks (for the first-inner edge channel), it is reasonable to model the MZI's visibility as a function of the FPI transmission, $T_{\text{FPI}} = \left| t^2 \sum_{n=0}^{\infty} (r^2 e^{i\delta \varphi_{FPI}})^n \right|^2 = \left| \frac{t^2}{1 - r^2 e^{i\delta \varphi_{FPI}}} \right|^2$, as $V_{\text{MZI}} = 1 - T_{\text{FPI}}$ [22]. For our simulation, we assumed a transmission $t = 0.75$.

Our model appears to apply at both $\nu_B = 2$ and $\nu_B = 3$, yet the observed behavior is different in the two cases. Apparently, the pairing physics, which is present only at $\nu_B = 3$, is essential. We do not know yet how to incorporate it into our model.

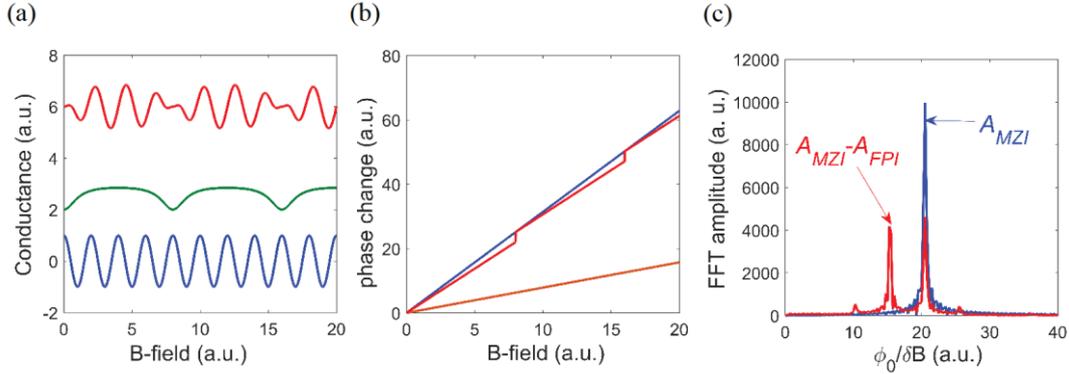

FIG. 16. Simulation for the conductance through the device shown in Fig. 7 in the main text. The device harbors an MZI and an FPI on one of its arms. We consider the configuration at $\nu_B = 3$ in which the MZI's QPCs partition the outermost edge channel (thus, inducing interference), and the FPI's QPCs fully transmit the outermost edge channel and fully reflect the two inner channels. (a) AB oscillations of an MZI with no FPI, for reference (blue), and of the MZI with the FPI (red) as functions of the magnetic field. As the magnetic field increases, in addition to the linear $A_{\text{MZI}} \delta B$ component, the MZI's phase accumulates an additional $\pi$ shift every time the first-inner channel (trapped inside the FPI) goes through a resonance (degeneracy between $N$ and $N + 1$, where $N$ is the number of electrons in the first-inner channel). This results in the dephasing of the MZI (red), as can be clearly seen from the visibility (green). (b) Phase evolution of the MZI without an FPI (blue) and with it (red). The brown plot (lowest plot) is the phase evolution of the FPI itself. (c) Fast Fourier transform of the AB oscillations shown in (a).